\documentclass[12pt,intlimits,reqno]{amsart}
\usepackage{amsmath}
\usepackage{enumerate}
\usepackage{amssymb}
\usepackage{amsthm}

\usepackage{bbm}

\usepackage{mathrsfs}

\usepackage[utf8]{inputenc}

\theoremstyle{definition}

\theoremstyle{remark}

\numberwithin{equation}{section}

\begin{document}

\title[Classical and quantum non-linear three-mode systems]{Integrability  and correspondence of classical and quantum non-linear three-mode systems}


\author{A. Odzijewicz and E. Wawreniuk}
\address{Institute of Mathematics\newline
University in Bia\l{}ystok\newline
Cio\l{}kowskiego~1M\newline
15-245 Bia\l{}ystok, Poland}
\email[A. Odzijewicz]{aodzijew@uwb.edu.pl}
\email[E. Wawreniuk]{ewawreniuk@math.uwb.edu.pl}


\begin{abstract}
The relationship between classical and quantum three one-mode systems interacting in a non-linear way is described. We investigate the integrability of these systems by using the reduction procedure. The reduced coherent states for the quantum system are constructed. We find the explicit formulas for the reproducing measure for these states. Examples of some applications of the obtained results in non-linear quantum optics are presented. 
 \end{abstract}

\maketitle
\tableofcontents

\section{Introduction}

Systems of three one-dimensional harmonic oscillators interacting in a non-linear way, see \eqref{pb} and \eqref{qH}-\eqref{comm}, supplies a capacious mathematical model for the description of various phenomena in mechanics as well as in non-linear optics. This concerns both classical and quantum cases. Usually, in order to avoid mathematical difficulties, the model is simplified by combining the classical and quantum approach. For example, one of modes is assumed to be in a coherent state which allows to treat it as a complex parameter. This reduces the three-mode system to the one which (being linear) is integrable. Also the evolution of this type of systems is usually considered in short time approximation which does not require information about the spectrum of Schr\"odinger operator. In order to stress the importance of three-mode non-linear systems let us mention a few most important physical phenomena modeled by it: sum-frequences and difference-frequences generations; Brillouin-Raman scattering which corresponds to the case when one of the modes is treated as a phonon; the interaction of one photon with the system of $N$ two-level atoms called superradiative emission.

This paper is partially based on \cite{KS}, where we have investigated on the classical and quantum level more general multi-mode systems. Applying methods of \cite{KS} here we study in detail the following setting. Finding appropriate canonical coordinates we reduce the classical system given by Hamiltonian \eqref{H} to a system \eqref{H5} of one-degree of freedom whose phase space is a cylinder. This allows one to solve the three-wave system of equations \eqref{twaveeq} in quadratures. Although results of Section \ref{sec1} are known, see e.g. \cite{armstrong}, they are important for the investigation presented in Section \ref{sec2}, where the correspondence between the classical and quantum systems is clarified. Namely, basing on the notion of standard (Glauber) coherent states we show that in the limit $\hbar \to 0$ the reduced canonical relations \eqref{comrel1}-\eqref{comrel3} correspond to Poisson algebra \eqref{pbx0xred2} and to the equation on the Kummer shape, see \eqref{casimir}.

In Section \ref{sec3} we show that after applying the quantum reduction procedure to the Hamiltonian operator \eqref{qH2} it splits on finite dimensional blocks. This allows us to obtain eigenvalues for arbitrary block of dimension not larger than nine. Examples of solution for the reduced Heisenberg equations are presented. We also present some applications of the obtained results in non-linear quantum optics.

In Section \ref{sec4}, applying the classical and quantum reduction procedure to standard coherent states we construct the reduced coherent states. The reproducing measure \eqref{rofun} for these states is obtained and the coherent state representation of the quantum reduced algebra \eqref{comrel1}-\eqref{comrel3} is given.

\section{Classical case and its integrability}\label{sec1}

In this section we will study the Hamiltonian system on   $\Omega^3 := \{ (z_0, z_1, z_2) \in \mathbb{C}^3 : |z_0|  >  0,   |z_1|  > 0,  |z_2| > 0 \}$,  equipped with the Poisson bracket 
\begin{equation}\label{pb}
\{f, g \} = - i \sum_{n=0}^2 \left(\frac{\partial f}{\partial z_n} \frac{\partial g}{\partial \bar z_n} -  \frac{\partial g}{\partial z_n} \frac{\partial f}{\partial \bar z_n}\right)
\end{equation}
of $f,g \in C^\infty (\Omega^3 )$ and Hamiltonian given by 
\begin{equation}\label{H}
H= \omega_0 |z_0|^2 + \omega_1 |z_1|^2 + \omega_2 |z_2|^2 + g_0 (z_0\bar{z}_1\bar{z}_2 + \bar{z}_0z_1z_2),
\end{equation}
where $g_0$ is a coupling constant and $\omega_0, \omega_1, \omega_2 $ are corresponding frequencies. The system of this type leads to the Hamilton equations:
\begin{equation}
\begin{aligned}
\frac{dz_0}{dt} &= i\omega_0 z_0 + ig_0 z_1z_2,\\
\label{twaveeq}
\frac{dz_1}{dt} & = i\omega_1 z_1 + ig_0 z_0 \bar{z}_2, \\
\frac{dz_2}{dt} & = i\omega_2 z_2 + i g_0 z_0 \bar{z}_1, 
\end{aligned}
\end{equation}
and is called a three-wave system \cite{Alber,Alber1}.  Other non-linear $n$-wave systems were integrated in \cite{gol1}.

Although the integrable Hamiltonian structure of the three-wave equations is well known, we will follow the general construction presented in \cite{KS} to reduce the considered system to a system of one degree of freedom. To this end we pass to the new canonical coordinates:
\begin{equation}\label{cor1}
\begin{array}{ccc}
I_0  := |z_0|^2, & I_1  := |z_0|^2 + |z_1|^2, & I_2 := |z_0|^2 +|z_2|^2,\\
\psi_0 := \phi_0-\phi_1-\phi_2, & \psi_1 :=\phi_1, & \psi_2:= \phi_2,
\end{array}
\end{equation}
where $z_k = |z_k|e^{i\phi_k}$ for $k=0,1,2$, i.e. the Poisson bracket \eqref{pb} written in these coordinates is given by 
\begin{equation}\label{pbcanonical}
\{ f, g \} = \sum_{n=0}^2 \left(\frac{\partial f}{\partial I_n} \frac{\partial g}{\partial  \psi_n} -  \frac{\partial g}{\partial I_n} \frac{\partial f}{\partial \psi_n}\right).
\end{equation}
From \eqref{cor1} we immediately obtain that:
\begin{equation}\label{ineq}
\begin{array}{ccc}
I_0  >  0  ,&  I_1 - I_0  > 0,& I_2-I_0 > 0,\\
0< \psi_1 \leq 2\pi,&  0< \psi_2 \leq 2\pi  , &  -4\pi < \psi_0 \leq 2\pi.
\end{array}
\end{equation}
The Hamiltonian flows generated by $I_0, I_1$ and $I_2$ are given by
\begin{equation}
\begin{aligned}
\sigma_0(t)(z_0, z_1, z_2) & = (e^{it}z_0, z_1, z_2), \\
 \sigma_1(t)(z_0,z_1,z_2) & = (e^{it}z_0, e^{it}z_1, z_2), \\
 \sigma_2(t)(z_0, z_1, z_2) & = (e^{it}z_0, z_1, e^{it}z_2),
\end{aligned} 
\end{equation}
respectively.


Hamiltonian \eqref{H} in coordinates $(I_0, I_1, I_2, \psi_0,\psi_1, \psi_2)$ takes the following form
\begin{equation}\label{H3}
H = (\omega_0 - \omega_1 - \omega_2) I_0 + \omega_1 I_1 + \omega_2 I_2 + 2g_0\sqrt{I_0(I_1-I_0)(I_2-I_0)} \cos \psi_0 . 
\end{equation}
Hence, one has three integrals of motion $H, I_1, I_2$ in involution and the integrals of motion $I_1$, $I_2$ can be considered as the components of the momentum map $\textbf{J}: \Omega^3 \to \mathbb{R}^2$, i.e.
\begin{equation}\label{mommap}
\textbf{J}(z_0, z_1, z_2) = \begin{pmatrix}
I_1 \\
I_2
\end{pmatrix} . 
\end{equation}
Note here that we identified $\mathbb{R}^2 $ with the dual of Lie algebra of the torus $\mathbb{T}^2:= \mathbb{S}^1 \times \mathbb{S}^1$. The map \eqref{mommap} is a submersion which maps $\Omega^3$ on $\mathbb{R}^2_+ := \{ (x_1, x_2) \in \mathbb{R}^2 : x_1 >0, x_2 > 0\}$. So, $\textbf{J}^{-1}(c_1, c_2)$ is a four-dimensional real submanifold of $\Omega^3$. Substituting $c_1, c_2$ instead of $I_1, I_2$ to inequalities  \eqref{ineq} we obtain that 
\begin{equation}\label{limitsi0}
0 < I_0 < c:=\min \{c_1, c_2\}
\end{equation}
and since coordinates $\psi_0, \psi_1, \psi_2$ are periodic we have $\textbf{J}^{-1} (c_1, c_2) \cong ]0, c [ \times \mathbb{S}^1 \times \mathbb{T}^2$. Hence, the reduced phase space $\textbf{J}^{-1} (c_1, c_2)/ \mathbb{T}^2 $  is isomorphic to the cylinder $]0, c [ \times \mathbb{S}^1$.

The variables $(I_0, \psi_0) \in ]0, c[\times \mathbb{S}^1$ form a canonical coordinates on the reduced phase space $]0, c[ \times \mathbb{S}^1$, i.e. 
\begin{equation}
\{I_0, \psi_0 \}_{\text{red}} = 1, \quad \{ \psi_0, \psi_0 \}_{\text{red}} = \{I_0, I_0\}_{\text{red}} =0. 
\end{equation}
So, the Hamiltonian \eqref{H3}  after reduction to $]0, c[ \times \mathbb{S}^1$ is given by 
\begin{equation}\label{H5}
H_{\text{red}} = (\omega_0 - \omega_1 - \omega_2) I_0 + \omega_1 c_1 + \omega_2 c_2 + 2g_0\sqrt{I_0(c_1-I_0)(c_2-I_0)} \cos \psi_0 . 
\end{equation}
The Hamilton equations for \eqref{H5} are 
\begin{align}
\label{pochizer}
\frac{d}{dt} I_0 &= 2 g_0 \sqrt{I_0 (c_1 - I_0)(c_2 - I_0)} \sin \psi_0, \\
\label{difpsizero}
\frac{d}{dt} \psi_0 &= (\omega_0 - \omega_1 - \omega_2) + g_0 \frac{3I_0^2-2(c_1 +c_2) I_0 + c_1c_2}{ \sqrt{I_0 (c_1 - I_0)(c_2 - I_0)}} \cos \psi_0. 
\end{align}
Since $H_{\text{red}} (I_0(t), \psi_0 (t) ) = E =\text{const.} $ one obtains from \eqref{H5} and \eqref{pochizer} differential equation on $I_0(t)$: 
\begin{equation}\label{difeqoni0}
\left(\frac{d}{dt}I_0(t)\right)^2 = 4 g_0^2 I_0(c_1-I_0)(c_2-I_0) - (E - (\omega_0 - \omega_1 - \omega_2 ) I_0 - \omega_1 c_1 - \omega_2 c_2 )^2 ,   
\end{equation}
which after separation of variables leads to the elliptic integral of the form
\begin{equation}\label{integral2}
t-t_0= \pm \int \frac{dI_0}{\sqrt{aI_0^3 +bI_0^2 +cI_0 +d}}, 
\end{equation}
where 
\begin{align}
a & := 4g_0^2,\\
b & := -4g_0^2(c_1+c_2) - (\omega_0 -\omega_1 - \omega_2)^2, \\
c & := 4g_0^2c_1c_2 +2H(\omega_0-\omega_1-\omega_2)- 2(\omega_0 - \omega_1 - \omega_2)(\omega_1c_1 + \omega_2c_2), \\
d & := 2H(\omega_1c_1 + \omega_2 c_2) - (\omega_1c_1 + \omega_2c_2)^2. 
\end{align} 
Substituting $ I_0 = \alpha s +\beta$, where \[
\alpha= g_0^{-\frac{2}{3}} \text{ and } \beta = -\frac{b}{3a}= \frac{4g_0^2(c_1+c_2)+(\omega_0-\omega_1 - \omega_2)^2}{12g_0^2},\]
 into elliptic integral on the right hand side of \eqref{integral2} we obtain the Weierstrass normal form of the elliptic integral of the first kind:
\begin{equation}\label{integral3}
g_0^{\frac{2}{3}}( t - t_0) = \int \frac{ds}{\sqrt{4s^3- g_2s- g_3}}, 
\end{equation}
where $g_2 = \alpha(\frac{b^2}{3a} -c)$ and $g_3 = -a\beta^3 -b\beta^2-c\beta -d$. From \eqref{integral3} and properties of the Weierstrass elliptic function $\wp $ \cite{akh} we obtain
\begin{equation}
 s (t)= \wp ( g_0^{\frac{2}{3}} (t-t_0) )
\end{equation}
and hence, 
\begin{equation}\label{izerowe}
I_0(t) = \alpha \wp ( g_0^{\frac{2}{3}}( t-t_0 ) )+ \beta . 
\end{equation}
Remembering that $I_1$ and $I_2$ are integrals of motion, one  can integrate equation \eqref{difpsizero} by quadratures obtaining time dependence of $\psi_0 (t)$. Next, having obtained $I_0(t)$ and $\psi_0 (t)$, one finds  $\psi_1 (t)$ and $\psi_2(t)$ integrating the equations
\begin{align}
\frac{d}{dt} \psi_1 & = \omega_1 + g_0 \sqrt{\frac{I_0(I_2-I_0)}{I_1 - I_0}}\cos \psi_0, \\
\frac{d}{dt} \psi_2 & = \omega_2 + g_0 \sqrt{\frac{I_0(I_1-I_0)}{I_2-I_0}} \cos \psi_0 , 
\end{align}
which are a part of the Hamilton equations given by Hamiltonian \eqref{H3}.

Another realization of the reduced phase space $\textbf{J}^{-1} (c_1, c_2)/ \mathbb{T}^2 $, called in \cite{Holm} the Kummer shape, can be obtained by providing a new complex variable 
\begin{equation}\label{variablezet}
z:= g_0 z_0 \bar z_1 \bar z_2 = g_0\sqrt{I_0(I_1-I_0)(I_2 -I_0)}e^{i\psi_0}. 
\end{equation}
Defining $x$ and $y$ as real and imaginary parts of $z=: x + i y$ one obtains a Poisson subalgebra 
\begin{align}\label{pbx0x}
\{I_0,x\} & =  - y,\\
 \{I_0,y\}& =   x,\\
\label{pbxy} 
\{x,y\} & =  \frac{1}{2} g_0^2 (3I_0^2 -2(I_1+I_2) I_0 + I_1I_2) ,\\
\{I_k , x \} & = \{I_k , y \} = 0 \mbox{ for } k=1,2,\\
\label{pbxykoniec}
\{ I_k, I_l\} & = 0, \mbox{ for } k,l=0,1,2,
\end{align}
of $(C^\infty(\Omega^3), \{ \cdot , \cdot \})$, generated by $I_0, I_1, I_2, x$ and $y$. 
Note that functions $I_0, x$ and $y$ are invariants of $\mathbb{T}^2$, so, they define the corresponding functions on the reduced phase space $\textbf{J}^{-1} (c_1, c_2)/ \mathbb{T}^2 $. Hence, we have a map of the reduced phase space $ ]0, c[\times \mathbb{S}^1 $ into $\mathbb{R}^3$ 
defined by 
\begin{equation}
\Phi_{c_1, c_2} (I_0, \psi_0) := \begin{pmatrix}
g_0 \sqrt{I_0(c_1-I_0)(c_2-I_0)}\cos \psi_0\\
 g_0 \sqrt{I_0(c_1-I_0)(c_2-I_0)}\sin \psi_0\\
 I_0 \end{pmatrix}. 
\end{equation}
This map is an embedding of $\textbf{J}^{-1}(c_1, c_2)/ \mathbb{T}^2 \cong ]0, c[ \times \mathbb{S}^1$ into $\mathbb{R}^3$. Its image is the $0$-level surface $\mathcal{C}^{-1}(0)$ of the function 
\begin{equation}\label{casimir}
\mathcal{C}(x,y,I_0) := - \frac{1}{2} ( x^2 + y^2 - g_0^2 I_0(c_1-I_0)(c_2-I_0))
\end{equation}
with removed points $\vec{0}= (0,0,0)$, $\vec{c} = (0,0,c)$. 
This bounded circularly symmetric surface $\mathcal{C}^{-1}(0) \backslash \{\vec{0}, \vec{c}\}$, which looks like a pinched sphere, is called Kummer shape \cite{Holm} or  the three-wave surface  \cite{Alber1}.

The functions $I_0 \circ \Phi_{c_1, c_2}, x\circ \Phi_{c_1, c_2}$ and $y\circ \Phi_{c_1, c_2}$, which are reduction of $I_0, x$ and $y$ to $\textbf{J}^{-1} (c_1, c_2)/ \mathbb{T}^2 $, satisfy the following relations
\begin{equation}\label{pbx0xred}
\begin{aligned}
\{I_0\circ \Phi_{c_1, c_2},x\circ \Phi_{c_1, c_2}\}_{\text{red}}  =&- y\circ \Phi_{c_1, c_2},\\
 \{I_0\circ \Phi_{c_1, c_2},y\circ \Phi_{c_1, c_2}\}_{\text{red}}  =& x\circ \Phi_{c_1, c_2},\\
\{x\circ\Phi_{c_1, c_2},y\circ \Phi_{c_1, c_2}\}_{\text{red}}  = &\frac{1}{2} g_0^2 (3(I_0\circ \Phi_{c_1, c_2})^2\\
& -2(c_1+c_2) (I_0\circ \Phi_{c_1, c_2}) + c_1c_2) .
\end{aligned}
\end{equation}
Taking the Poisson algebra of smooth functions  $(C^\infty (\mathbb{R}^3), \{ \cdot , \cdot \}_\mathcal{C} )$  with Nambu-Poisson bracket 
\begin{equation}\label{nambu}
\{f,g\}_\mathcal{C}:=\det [\nabla \mathcal{C},\nabla f, \nabla g],
\end{equation}
 of $f,g \in C^\infty (\mathbb{R}^3)$, where function $\mathcal{C}$ is defined in \eqref{casimir} and
where $ \nabla f = \left( \frac{\partial f}{\partial x} , \frac{\partial f}{\partial y}, \frac{ \partial f}{\partial I_0 } \right)^T$, we find that coordinate functions $I_0, x,y \in C^\infty (\mathbb{R}^3)$ satisfy
\begin{equation}\label{pbx0xred2}
\begin{aligned}
\{I_0,x\}_{\mathcal{C}} & =- y,\\
 \{I_0,y\}_{\mathcal{C}} & = x,\\
\{x,y\}_{\mathcal{C}} & = \frac{1}{2} g_0^2 (3I_0^2 -2(c_1+c_2) I_0 + c_1c_2) .
\end{aligned}
\end{equation}
It follows from \eqref{pbx0xred} and \eqref{pbx0xred2} that $\Phi_{c_1, c_2} : ]0, c[\times \mathbb{S}^1 \to \mathbb{R}^2 \times ]0, c[$ is a Poisson map, which maps the reduced phase space $\textbf{J}^{-1}(c_1, c_2)/\textbf{T}^2 \cong ]0, c[ \times \mathbb{S}^1 $ on the symplectic leaf $\mathcal{C}^{-1}(0)\backslash \{\vec{0}, \vec{c} \}$ of the Poisson algebra $(C^\infty(\mathbb{R}^3 ), \{ \cdot , \cdot \}_{\mathcal{C}})$. Note here that $ \mathcal{C}: \mathbb{R}^2 \times ]0, c[ \to \mathbb{R}$ defined in \eqref{casimir} is a Casimir function for  $(C^\infty (\mathbb{R}^3) , \{ \cdot , \cdot \}_{\mathcal{C}})$. Consequently, the circularly symmetric surfaces $\mathcal{C}^{-1}(\lambda ) $, $\lambda \in \mathbb{R}$, are symplectic leaves for the considered case. 

The dynamics given by Hamiltonian 
\begin{equation}\label{H4}
H_{\mathcal{C}} := (\omega_0 - \omega_1 - \omega_2) I_0 + \omega_1 c_1 + \omega_2 c_2 + 2x. 
\end{equation}
on $(\mathbb{R}^2 \times ]0, c[, \{ \cdot , \cdot \}_\mathcal{C})$, after reduction to the symplectic leaf $\mathcal{C}^{-1}(0)\backslash \{\vec{0}, \vec{c}\}$ coincides with the one obtained for Hamiltonian \eqref{H5}. 
The Hamilton equations for this Hamiltonian defined by the Nambu-Poisson bracket \eqref{nambu} are
\begin{align}
\label{2igrek}
 \frac{dI_0}{dt} & = \{H_{\mathcal{C}}, I_0\}_\mathcal{C}= 2y, \\
\frac{dx}{dt} & = \{H_{\mathcal{C}}, x\}_\mathcal{C}=-(\omega_0 -\omega_1 - \omega_2 )y, \\
\label{2igrek2}
\frac{dy}{dt} & = \{ H_{\mathcal{C}}, y \}_\mathcal{C} = (\omega_0 - \omega_1 - \omega_2)x + g_0^2 (3I_0^2-2(c_1 +c_2) I_0 + c_1c_2).
\end{align}
Hence, we obtain 
\begin{gather}\label{iksowe}
x(t) = \frac{1}{2}(E - (\omega_0 - \omega_1 - \omega_2)I_0(t) - \omega_1 c_1 - \omega_2 c_2),\\
\label{igrekowe}
y(t) = \frac{1}{2} \frac{dI_0}{dt}(t), 
\end{gather}
where $E$ is a fixed constant (total energy of the system). 
Substituting \eqref{iksowe} and \eqref{igrekowe} into equation \eqref{2igrek2} we obtain the second order differential equation on $I_0(t)$
\begin{multline}\label{difeq2}
\frac{d^2}{dt^2} I_0(t) = (\omega_0- \omega_1 - \omega_2)(E - (\omega_0 - \omega_1 - \omega_2)I_0 - \omega_1 c_1 - \omega_2 c_2)  \\
+2g_0^2 (3I_0^2-2(c_1 +c_2) I_0 + c_1c_2),
\end{multline}
which solution coincides (up to a constant) with the one obtained from equation \eqref{difeqoni0}. Concluding, let us mention that one can obtain the trajectory of $t \mapsto (x(t), y(t), I_0(t)) \in \mathcal{C}^{-1}(0)\backslash \{\vec{0}, \vec{c}\}$ defined by \eqref{2igrek}-\eqref{2igrek2} as an intersection of the Kummer shape $\mathcal{C}^{-1}(0)\backslash \{\vec{0}, \vec{c}\}$ with the hyperplane given by the level set $H_\mathcal{C} ^{-1}(E)$ of the Hamiltonian \eqref{H4}.

\section{Correspondence between quantum and classical systems}\label{sec2}

Quantum counterpart of the classical hamiltonian system considered in Section \ref{sec1} is given by the following three-boson Hamiltonian
\begin{equation}\label{qH}
\hat{H} = \omega_0 a^*_0a_0 + \omega_1 a_1^*a_1 + \omega_2 a_2^*a_2 + g_0 (a_0a_1^*a_2^* + a_0^*a_1a_2), 
\end{equation}
expressed by creation and annihilation operators satisfying the canonical commutation relations
\begin{equation}\label{comm}
[a_i, a_j^*] = \hbar \delta_{ij}, \qquad [a_i, a_j] = 0, \qquad [a_i^*, a_j^*] = 0, 
\end{equation}
where $\hbar$ is Planck constant and real parameters $\omega_0, \omega_1, \omega_3$ and $g_0$ have the same physical meaning as in \eqref{H}.


In order to integrate the quantum system \eqref{qH}, similarly to the classical case, one introduces new quantum coordinates:
\begin{equation}\label{defa0}
\begin{aligned}
A &:= g_0 a_0a_1^*a_2^*, \\
A^* &:= g_0 a_0^* a_1 a_2,\\
A_0 &:= a_0^*a_0, \\
A_1 & := a_0^*a_0 +a_1^*a_1, \\
A_2 & := a_0^*a_0 +a_2^*a_2,  
\end{aligned}
\end{equation}
see \cite{KS, AO2}. 
The Hamiltonian \eqref{qH} expressed in terms of $A_0, A_1, A_2, A$ and $A^*$ takes the form
\begin{equation}\label{qH2}
\hat{H} = (\omega_0-\omega_1 - \omega_2)A_0 + \omega_1 A_1 + \omega_2 A_2 +A +A^*. 
\end{equation}

The operators $A_0, A_1, A_2$ are diagonal in the Fock basis
\begin{equation}\label{fock}
|n_0, n_1, n_2 \rangle := \frac{1}{\sqrt{n_0! n_1! n_2!}} \hbar^{-\frac{1}{2}(n_0 +n_1 + n_2)} (a_0^*)^{n_0} (a_1^*)^{n_1} (a_2^*)^{n_2} |0, 0, 0\rangle , 
\end{equation}
where $n_0, n_1, n_2 \in \mathbb{N}\cup \{0\}$ and $|0, 0, 0\rangle $ is the vacuum vector, i.e. $a_i |0, 0 , 0\rangle = 0 $ for $i=0,1,2$. Notice that the eigenvalues $c_0, c_1, c_2$ of $A_0, A_1, A_2$ with the corresponding eigenvectors $|n_0, n_1, n_2\rangle $ are given by 
\begin{equation}
c_0 = \hbar n_0, \quad c_1 = \hbar (n_0 + n_1 ), \quad c_2 = \hbar (n_0 + n_2). 
\end{equation}
Obviously operators $A_0, A_1, A_2, A$ and $A^*$ satisfy the relations
\begin{gather}\label{comrelf1}
[A_0, A] = -\hbar A , \quad [A_0, A^*] = \hbar A^*, \\
\label{comrelf}
AA^* = g^2_0 (A_0 + \hbar) (A_1 - A_0)(A_2-A_0), \\
A^*A = g_0^2 A_0 (A_1 - A_0 +\hbar )(A_2-A_0 + \hbar ). \\
[A_k , A] = [A_k, A^*] =0 \mbox{ for } k=1,2,\\
\label{comrelf2}
[A_k , A_l ]= 0 , \mbox{ for } k,l=0,1,2, \quad  
\end{gather}
Using the above relations we find that Heisenberg equations on $A_0(t), A_1 (t), A_2(t), X(t)$ and $Y(t)$ for the Hamiltonian \eqref{qH2} are:
\begin{align}
\label{hea01}\frac{d}{dt} A_0 (t) = &2 Y(t), \\
\label{calka1}
\frac{d}{dt}  A_1 (t) = &0, \\
\label{calka2}
\frac{d}{dt}  A_2 (t)  =& 0, \\
\frac{d}{dt}  X (t)  = &-(\omega_0 -\omega_1 -\omega_2)  Y (t), \\
\nonumber\frac{d}{dt}  Y (t)  = &(\omega_0 - \omega_1 -\omega_2)  X(t) + \\
\nonumber  &+g_0^2(3 A_0^2(t) - 2 ( A_1 + A_2 )  A_0 (t) +\\
\label{heyt1}  &+ A_1  A_2  - \hbar  A_0(t) ), 
\end{align}
where $X(t) := \frac{1}{2} (A(t) + A^*(t))$, $Y(t) := \frac{1}{2i} (A(t) - A^*(t))$.

Now let us discuss the correspondence between the classical and quantum systems. In this context the notion of the standard coherent states is crucial, which for a three-mode system are:
\begin{equation}\label{Glauber}
|z_0, z_1, z_2 \rangle := \sum_{n_0, n_1, n_2 =0}^\infty \frac{z_0^{n_0}z_1^{n_1} z_2^{n_2}}{\sqrt{n_0! n_1! n_2! }} \hbar^{-\frac{1}{2}(n_0+n_1+n_2)} |n_0, n_1, n_2 \rangle ,
\end{equation} 
where $z_0, z_1, z_2 \in \mathbb{C}$ and $|n_0, n_1, n_2 \rangle $ is an element of the Fock basis given by \eqref{fock}. Note that coherent states \eqref{Glauber} are the eigenvectors of the annihilation operators $a_0, a_1$ and $a_2$, i.e. 
\begin{equation}\label{proste1}
a_i |z_0, z_1, z_2 \rangle = z_i |z_0, z_1, z_2 \rangle \mbox{ for  } i=0,1,2,. 
\end{equation}
We have also the resolution of identity for the standard coherent states:
\begin{equation}\label{resstandard}
\mathbbm{1} = \int_{\Omega^3} \frac{|w_0, w_1, w_2 \rangle \langle w_0, w_1, w_2 |}{\langle w_0, w_1, w_2 | w_0, w_1, w_2 \rangle } d\mu_\hbar (\bar w_0, \bar w_1 , \bar w_2, w_0, w_1, w_2 ), 
\end{equation}
where
\[ 
d\mu_\hbar (\bar w_0, \bar w_1 , \bar w_2, w_0, w_1, w_2 ) = \frac{1}{(2i\pi \hbar )^3} d w_0 \wedge d\bar w_0  \wedge d w_1 \wedge d\bar w_1 \wedge d w_2 \wedge d\bar w_2\]
is the Liouville measure appropriately normalized.

Let us define the covariant symbol $\langle F \rangle_\hbar $ of an operator $F$ as its mean value 
\begin{equation}\label{mvG}
\langle F \rangle_\hbar  (\bar z_0 , \bar z_1, \bar z_2, z_0, z_1, z_2) := \frac{\langle z_0, z_1, z_2 |F|z_0, z_1 z_2 \rangle }{\langle z_0, z_1, z_2 | z_0, z_1, z_2 \rangle }
\end{equation}
on the coherent states \eqref{Glauber}. Since there is one-to-one correspondence between operators and their symbols one can define the $*_\hbar $-product of covariant symbols of operators $F$ and $G$ in the following way
\begin{multline}\label{productzg}
(\langle F \rangle_\hbar *_\hbar \langle G \rangle_\hbar ) (\bar z_0 , \bar z_1, \bar z_2, z_0, z_1, z_2) := \frac{\langle z_0, z_1, z_2 |FG|z_0, z_1 z_2 \rangle }{\langle z_0, z_1, z_2 | z_0, z_1, z_2 \rangle }\\
= \int_{\Omega^3} \langle F \rangle_\hbar  (\bar z_0 , \bar z_1, \bar z_2, w_0, w_1, w_2) \langle G \rangle_\hbar  (\bar w_0 , \bar w_1, \bar w_2, z_0, z_1, z_2)\\
\times e^{-\frac{1}{\hbar}(|z_0 - w_0|^2 + |z_1 - w_1 |^2 + |z_2 - w_2 |^2)} d\mu_\hbar (\bar w_0, \bar w_1 , \bar w_2, w_0, w_1, w_2 ) , 
\end{multline}
where the second equality follows from the resolution of identity \eqref{resstandard}. The above definition of the $*_\hbar $-product is equivalent to the standard one
\begin{multline}\label{standsp}
(\langle F \rangle_\hbar *_\hbar \langle G \rangle_\hbar ) (\bar z_0 , \bar z_1, \bar z_2 , z_0, z_1, z_2 )  \\
 := \sum_{j_0, j_1 , j_2 =0}^\infty \frac{\hbar^{j_0 + j_1 + j_2 }}{j_0! j_1! j_2!} \frac{\partial^{j_0+j_1+j_2}}{\partial z_0^{j_0}\partial z_1^{j_1}\partial z_2^{j_2}} \langle F \rangle_\hbar (\bar z_0, \bar z_1 , \bar z_2 , z_0, z_1 , z_2 )  \\
 \times \frac{\partial^{j_0+j_1+j_2}}{\partial \bar z_0^{j_0}\partial \bar z_1^{j_1}\partial \bar z_2^{j_2}} \langle G \rangle_\hbar (\bar z_0, \bar z_1 , \bar z_2 , z_0, z_1 , z_2 ).
\end{multline}
Let us stress that the dependence on $\hbar$ of the covariant symbol \eqref{mvG} and the star product \eqref{productzg}  is given through the coherent states \eqref{Glauber}. According to \eqref{standsp} one has
\begin{equation}\label{limitsp}
\langle F \rangle_\hbar *_\hbar \langle G \rangle_\hbar \underset{\hbar \to 0 }{\longrightarrow } \langle F \rangle \cdot \langle G \rangle 
\end{equation}
 and 
\begin{equation}\label{310}
\lim_{\hbar \to 0 } \frac{-i}{\hbar} (\langle F \rangle_\hbar *_\hbar \langle G \rangle_\hbar- \langle G \rangle_\hbar *_\hbar \langle F \rangle_\hbar )  = \{ \langle F \rangle, \langle G \rangle \},
\end{equation}
where $\lim_{\hbar \to 0} \langle F\rangle_\hbar = \langle F \rangle$.

Computing the covariant symbols of operators $A_0, A_1, A_2, A$ and $A^*$ one obtains
\begin{gather}\label{relacjeqc}
\langle A_0 \rangle_\hbar = I_0, \quad \langle A_1 \rangle_\hbar = I_1 , \quad \langle A_2 \rangle_\hbar = I_2, \\
\label{relacjeqc2}
\langle A \rangle_\hbar = z, \quad \langle A^* \rangle_\hbar = \bar z .
\end{gather}
Note that the above covariant symbols do not depend on $\hbar$. 
From this fact and from the relations \eqref{comrelf1}-\eqref{comrelf2} one finds the relations corresponding to the covariant symbols \eqref{relacjeqc}-\eqref{relacjeqc2}:
\begin{gather}\label{csrel2}
I_0 *_\hbar z -  z *_\hbar I_0= -\hbar z,\\
I_0 *_\hbar \bar z -  \bar z *_\hbar I_0= \hbar \bar z,\\
z *_\hbar \bar z -  \bar z *_\hbar z=  \hbar g_0^2 (3I_0^2 -2(I_1+I_2)I_0 +I_1I_2 - \hbar I_0).\\
I_k *_\hbar z -  z *_\hbar I_k = I_k *_\hbar \bar z -  \bar z *_\hbar I_k =0, \mbox{ for } k=1,2, \\
\label{csrel1}
I_k *_\hbar I_l -  I_l *_\hbar I_k = 0 \mbox{ for } k,l=0,1,2, 
\end{gather}
Taking the above into account, from \eqref{310} one sees that in the classical limit $\hbar \to 0$ the relations \eqref{pbx0x}- \eqref{pbxykoniec} give exactly the ones  for the classical counterparts of $A_0, A_1, A_2, X$ and $Y$. Therefore, the Poisson algebra defined by \eqref{pbx0x}- \eqref{pbxykoniec} is the classical limit of the quantum algebra defined by the relations \eqref{comrelf1}-\eqref{comrelf2}.

Computing the covariant symbols of both sides of the Heisenberg equations \eqref{hea01}-\eqref{heyt1} on the operators $A_0, A_1,A_2 , X$ and $Y$ and taking the fact that $A_1$, $A_2$ are integrals of motion into account, one obtains the following equations 
 \begin{align}
\label{hea0}\frac{d}{dt} \langle A_0 (t)\rangle_\hbar = &2 \langle Y(t)\rangle_\hbar, \\
\frac{d}{dt} \langle A_1 (t)\rangle_\hbar =& 0, \\
\frac{d}{dt} \langle A_2 (t) \rangle_\hbar = &0, \\
\frac{d}{dt} \langle X (t) \rangle_\hbar = &-(\omega_0 -\omega_1 -\omega_2) \langle Y (t)\rangle_\hbar, \\
\nonumber\frac{d}{dt} \langle Y (t)\rangle_\hbar  = &(\omega_0 - \omega_1 -\omega_2) \langle X(t)\rangle_\hbar  \\
\nonumber  &+g_0^2(3\langle A_0(t)\rangle_\hbar *_\hbar \langle A_0(t) \rangle_\hbar - 2 (\langle A_1\rangle_\hbar +\langle A_2 \rangle_\hbar) *_\hbar\langle A_0 (t)\rangle_\hbar \\
\label{heyt}  &+\langle A_1 \rangle_\hbar *_\hbar \langle A_2 \rangle_\hbar - \hbar \langle A_0(t)\rangle_\hbar ). 
\end{align}
on the covariant symbols $\langle A_0(t) \rangle_\hbar$, $\langle A_1(t) \rangle_\hbar$, $\langle A_2(t) \rangle_\hbar$, $\langle X(t) \rangle_\hbar$, $\langle Y(t) \rangle_\hbar$. Let us note that the covariant symbols of $A_1(t)$ and $A_2(t)$ do not depend on time, hence $\langle A_1 (t)\rangle_\hbar = \langle A_1 \rangle_\hbar = I_1$ and $\langle A_2 (t)\rangle_\hbar = \langle A_2 \rangle_\hbar= I_2$. 
From equation \eqref{qH2} one has
\begin{equation}\label{qxt}
\langle X(t) \rangle_\hbar = \frac{1}{2} (\langle H \rangle_\hbar - (\omega_0 -\omega_1 - \omega_2)\langle A_0(t) \rangle_\hbar- \omega_1 I_1 - \omega_2 I_2).
\end{equation}
Substituting \eqref{hea0} and \eqref{qxt} to \eqref{heyt} one obtains the equation on the covariant symbol $\langle A_0 (t) \rangle_\hbar $:
\begin{multline}\label{eqcsa0}
\frac{d^2}{dt^2} \langle A_0(t) \rangle_\hbar = \\
=6g_0^2 \sum_{j_0, j_1 , j_2 =0}^\infty \frac{\hbar^{j_0 + j_1 + j_2 }}{j_0! j_1! j_2!} \left(\frac{\partial^{j_0}}{\partial z_0^{j_0}} \frac{\partial^{j_1}}{\partial z_1^{j_1}} \frac{\partial^{j_2}}{\partial z_2^{j_2}}\right) \langle A_0 (t) \rangle_\hbar  \left(\frac{\partial^{j_0}}{\partial \bar z_0^{j_0}} \frac{\partial^{j_1}}{\partial \bar z_1^{j_1}} \frac{\partial^{j_2}}{\partial \bar z_2^{j_2}}\right)\\ \times\langle A_0 (t) \rangle_\hbar 
 -(4g_0^2(I_1+I_2 +\hbar (2\bar z_0 \frac{\partial}{\partial \bar z_0} + \bar z_1 \frac{\partial}{\partial \bar z_1} + \bar z_2 \frac{\partial}{\partial \bar z_2})) \\
+2g_0^2\hbar + (\omega_0 - \omega_1 - \omega_2)^2) \langle A_0(t) \rangle_\hbar 
+(\omega_0 -\omega_1-\omega_2)(\langle H \rangle_\hbar - \omega_1 I_1 - \omega_2 I_2) \\
+ 2g_0^2 (I_1I_2 +\hbar I_0). 
\end{multline} 
Note here that $I_1, I_2$ and $\langle H \rangle_\hbar$ do not depend on $t$. In the limit $\hbar \to 0$ the above equation gives 
\begin{multline}\label{eqcsa01}
\frac{d^2}{dt^2} \langle A_0(t) \rangle = (\omega_0 - \omega_1 - \omega_2)(E - \omega_1 I_1 - \omega_2 I_2) \\
+2g_0^2 (3 \langle A_0(t) \rangle^2- 2(I_1 + I_2)\langle A_0(t)\rangle+ I_1 I_2) -(\omega_0 -\omega_1 - \omega_2)^2 \langle A_0(t)\rangle,
\end{multline}
where $\langle A_0(t)\rangle := \lim_{\hbar \to 0} \langle A_0(t) \rangle_\hbar$ and $E:= \lim_{\hbar \to 0}\langle H \rangle_\hbar$. 
Reducing this equation to $\textbf{J}^{-1}(c_1, c_2) / \mathbb{T}^2$ one must put $c_1, c_2$ in \eqref{eqcsa01}  instead of $I_1, I_2$, which  gives the differential equation \eqref{difeq2} on $I_0(t)$ obtained in Section~2. Since additionally $\langle A_0 \rangle_\hbar = I_0$,  we get that in the classical limit $\hbar \to 0$ the covariant symbol of $A_0(t)$ is exactly the function $I_0(t)$. From equations \eqref{qxt},\eqref{hea0} one obtains equations \eqref{iksowe} and \eqref{igrekowe} on the classical counterparts $x(t), y(t)$ of $X(t)$, $Y(t)$.

Summarizing, we state that in  the classical limit $\hbar \to 0$ considered here the quantum system corresponds to the classical one which was described in the previous section.

\section{Some remarks about the integrability and applications of the quantum case}\label{sec3}
Having two commuting quantum integrals of motion $A_1, A_2$, see \eqref{calka1}-\eqref{calka2} and \eqref{comrelf2}, we can reduce the quantum system described by the Hamiltonian \eqref{qH} to the Hilbert subspaces $\mathcal{H}_{c_1, c_2}$, $c_1, c_2 \in \hbar\mathbb{Z}_+:=\{ n\in \mathbb{Z}: n\geq 0\}$,  spanned by the common eigenvectors of $A_0, A_1, A_2$ with fixed eigenvalues $c_1, c_2$ of $A_1, A_2$. The dimension of the subspace $\mathcal{H}_{c_1, c_2}$ is  $\dim_\mathbb{C} \mathcal{H}_{c_1, c_2} = L +1$, where $L:= \min \{v_1, v_2 \}$, and the set of vectors 
\begin{equation}\label{basishc}
\{ |n, v_1 -n, v_2 -n \rangle \}_{n=0}^L 
\end{equation}
forms a basis of $\mathcal{H}_{c_1, c_2}$. 

After reduction of $A$, $A^*$ and $A_0$ to $\mathcal{H}_{c_1, c_2} $ the reduced operators $\textbf{A}, \textbf{A}^*$ and $\textbf{A}_0$ in the basis \eqref{basishc} are given by 
\begin{equation}\label{a0nabaze}
\textbf{A}_0 |n, v_1 -n, v_2 -n \rangle  = \hbar n |n, v_1-n, v_2-n \rangle , 
\end{equation}
\begin{multline}\label{anabaze}
\textbf{A} |n, v_1-n, v_2-n\rangle  
=g_0 \sqrt{\hbar^3 n(v_1-n+1)(v_2-n+1)} \\
\times|n-1, v_1-n+1, v_2-n+1 \rangle  , 
\end{multline}
\begin{multline}\label{azgnabaze}
\textbf{A}^* |n, v_1-n, v_2 -n\rangle 
= g_0\sqrt{\hbar^3 (n+1) (v_1-n)(v_2-n)}\\
\times|n+1, v_1-n-1, v_2-n-1 \rangle .
\end{multline}
It follows from \eqref{anabaze} that $\textbf{A}|0,v_1,v_2\rangle =0$, $v_1, v_2 \in \mathbb{N}\cup \{0\}$,   and $\textbf{A}$ and $\textbf{A}^*$ are lowering and raising  operators, which satisfy $\textbf{A}|0,v_1,v_2\rangle =0$ and $\textbf{A}^* |L, v_1 - L, v_2 - L \rangle = 0$. Acting on $|0, v_1, v_2\rangle$ by $\textbf{A}^*$ one generates the basis \eqref{basishc}. So, the vector $|0, v_1, v_2 \rangle $ can be considered as the vacuum state for the reduced quantum system described by the following relations 
\begin{gather}\label{comrel1}
[\textbf{A}_0, \textbf{A}] = -\hbar \textbf{A}, \quad [\textbf{A}_0, \textbf{A}^*] = \hbar \textbf{A}^*, \\
\label{comrel2}
\textbf{A}\textbf{A}^* = g_0^2 (\textbf{A}_0+\hbar) (\hbar v_1 - \textbf{A}_0  )(\hbar v_2-\textbf{A}_0  ), \\
\label{comrel3}
\textbf{A}^*\textbf{A} = g_0^2 \textbf{A}_0 (\hbar v_1 - \textbf{A}_0 +\hbar )(\hbar v_2-\textbf{A}_0 + \hbar ).
\end{gather}
Note that operator algebra generated by reduced operators $\textbf{A}_0, \textbf{A}$ and $\textbf{A}^*$ is  the algebra of $(L+1)\times (L+1)$ matrices over $\mathbb{C}$. Replacing in \eqref{comrel1}-\eqref{comrel3} $\textbf{A}$ and $\textbf{A}^*$ by $\textbf{X} := \frac{1}{2}(\textbf{A} + \textbf{A}^*)$ and $\textbf{Y} := \frac{1}{2i} (\textbf{A} - \textbf{A}^*)$ we obtain the relations
\begin{equation}\label{qpbx0xred2}
\begin{aligned}[]
[\textbf{A}_0, \textbf{X}]
=& -\hbar  i \textbf{Y}, \\
[\textbf{A}_0, \textbf{Y}] =& \hbar i \textbf{X},\\
[\textbf{X}, \textbf{Y} ] =& \hbar (3\textbf{A}_0^2 - 2\hbar(v_1+v_2)\textbf{A}_0 +\hbar^2 v_1v_2 - \hbar \textbf{A}_0),
\end{aligned}
\end{equation}
which one can interpret as the quantum variant of the relations \eqref{pbx0xred2} defining the Poisson algebra $(C^\infty (\mathbb{R}^3), \{ \cdot, \cdot \}_\mathcal{C})$. The relation
\begin{equation}
\textbf{X}^2 + \textbf{Y}^2 = \frac{1}{2} g_0^2 (2 \textbf{A}_0^3 - (2\hbar(v_1+v_2) +\hbar)\textbf{A}_0^2 +(2\hbar^2v_1v_2 + \hbar^2)\textbf{A}_0 + \hbar^3v_1v_2),
\end{equation}
is a quantum counterpart of the Kummer shape $\mathcal{C}^{-1}(0)\backslash \{\vec{0}, \vec{c} \}$ defined by the function \eqref{casimir}. 

The Hamiltonian \eqref{qH2} after reduction to $\mathcal{H}_{c_1, c_2}$ is given by 
\begin{multline}\label{redqH}
\hat{\textbf{H}} = (\omega_0 - \omega_1 - \omega_2)\textbf{A}_0 -\omega_1 c_1 - \omega_2 c_2 + \textbf{A}+\textbf{A}^*\\
= (\omega_0 - \omega_1 - \omega_2)\textbf{A}_0 -\omega_1 c_1 - \omega_2 c_2 + 2\textbf{X}. 
\end{multline}
Therefore, the Heisenberg equations for  $\textbf{A}_0$, $\textbf{X}(t)$ and $\textbf{Y}(t) $ are the following
\begin{align}\label{qa0t}
\frac{d}{dt} \textbf{A}_0 (t) =& 2 \textbf{Y}(t), \\
\label{qheixt}
\frac{d}{dt} \textbf{X} (t) = &-(\omega_0 -\omega_1 -\omega_2) \textbf{Y} (t), \\
\label{qyt}
\frac{d}{dt} \textbf{Y} (t) =& (\omega_0 - \omega_1 -\omega_2) \textbf{X}(t) + g_0^2(3\textbf{A}_0^2 (t)\nonumber\\
&- 2 (c_1 +c_2) \textbf{A}_0 (t) +c_1c_2 - \hbar \textbf{A}_0(t)). 
\end{align}
From \eqref{redqH} and \eqref{qa0t} we 
obtain the equation
\begin{multline}\label{opeqa0}
\frac{d^2}{dt^2} \textbf{A}_0 (t) = 6g_0^2 \textbf{A}_0^2(t) -(4g_0^2(c_1+c_2) +2\hbar +(\omega_0-\omega_1-\omega_2)^2) \textbf{A}_0(t) \\
+2g_0^2c_1c_2 +(\omega_0-\omega_1-\omega_2)(\hat{\textbf{H}} - \omega_1c_1 -\omega_2c_2)
\end{multline}
for $\textbf{A}_0(t)$. In order to obtain the solution $\textbf{A}_0(t) = e^{-\frac{i}{\hbar}t\hat{\textbf{H}}}\textbf{A}_0 e^{\frac{i}{\hbar}t\hat{\textbf{H}}}$ 
of \eqref{opeqa0}, we need to find the eigenvalues of the reduced Hamiltonian $\hat{\textbf{H}}$.

In what follows we assume the condition $\omega_0 - \omega_1 - \omega_2 =0$ for frequencies $\omega_0, \omega_1$ and $\omega_2$, which in various physical models of quantum optics is a consequence of the energy conservation law. Notice that under this condition $\textbf{X}$ 
is an integral of motion.

The Hamiltonian \eqref{redqH} in the basis \eqref{basishc} is a three-diagonal matrix, which can be written in the following way:
\begin{equation}\label{matqH}
\hat{\textbf{H}} = (\omega_1 c_1 + \omega_2 c_2) \mathbbm{1}_{L+1} + \hat{\textbf{H}}_I,
\end{equation}
where $\mathbbm{1}_{L+1}$ is $(L+1)\times (L+1)$ identity matrix and 
\begin{equation}\label{qh1}
\hat{\textbf{H}}_I = \begin{pmatrix}
0 & b_1 & 0 & \ldots & 0&0 \\
b_1 & 0 &b_2 & \ldots & 0&0 \\
0 & b_2 &0&\ldots &0&0\\
\vdots& \vdots & \vdots & \ddots & \vdots & \vdots \\
0&0&0& \ldots &0 & b_L\\
0&0&0& \ldots &b_L &0
\end{pmatrix} ,
\end{equation}
where the coefficients of $\vec{b}_L := (b_1, b_2, \ldots , b_L)^T$ are defined by
\begin{equation}\label{beka}
b_k :=  g_0 \sqrt{\hbar^3 k(v_1-k+1)(v_2-k+1)}.
\end{equation}

Denote by $\Delta_k (\lambda )$ the leading principal minor of $\hat{\textbf{H}}_I - \lambda \mathbbm{1}_{L+1}$ of order $k$, i.e. 
\begin{equation}\label{defdeltak}
\Delta_k (\lambda ) := \det \begin{pmatrix}
-\lambda & b_1 & \ldots & 0&0 \\
b_1 & -\lambda & \ldots & 0&0 \\
\vdots& \vdots &  \ddots & \vdots & \vdots \\
0&0& \ldots &-\lambda & b_{k-1}\\
0&0& \ldots &b_{k-1} &-\lambda
\end{pmatrix},
\end{equation}
for $k=2, \ldots, L+1$, $\Delta_1 (\lambda) = -\lambda$ and by definition  $\Delta_0 (\lambda ):=1$. From the Laplace expansion of \eqref{defdeltak} one obtains the three-term recurrence formula 
\begin{equation}\label{recfdeltak}
\Delta_k (\lambda) = - \lambda \Delta_{k-1} (\lambda ) - b_{k-1}^2 \Delta_{k-2} (\lambda), 
\end{equation} 
where $ k=2,3,\ldots L+1$.

Since $\Delta_0(\lambda) $ is an even polynomial and $\Delta_1 (\lambda )$ is an odd polynomial, we have 
 $\Delta_k (\lambda ) = \Delta_k (-\lambda )$ for even $k$ and $\Delta_k (\lambda) = - \Delta_k (-\lambda ) $ for odd $k$. Thus, for the characteristic polynomial $\Delta_{L+1} (\lambda) = \det (\hat{\textbf{H}}_I - \lambda \mathbbm{1}_{L+1})$ of $\hat{\textbf{H}}_I$ we have
\begin{equation}
\begin{aligned}\label{proposition}
  & \Delta_{L+1}(\lambda) = \lambda \Omega_K (\lambda^2 )&&\text{if } L=2K,\\
  & \Delta_{L+1}(\lambda ) = \Theta_K (\lambda^2 ) &&\text{if } L= 2K-1 , 
\end{aligned}
\end{equation}
where $\Omega_K$ and $\Theta_K$ are polynomials of degree $K\in \mathbb{N}$.

Notice that formula \eqref{recfdeltak} allows us to easily compute the characteristic polynomial of $\hat{\textbf{H}}_I$ for any dimension of $\mathcal{H}_{c_1, c_2}$. For example,  the characteristic polynomials \eqref{proposition} up to 5-th dimension of $\mathcal{H}_{c_1, c_2}$ are given by
\begin{equation}\label{tabela4}
\begin{aligned}
\Delta_2(\lambda )  =& \lambda^2 -\vec{b}_1^2= (\lambda - b_1)(\lambda+ b_1),\\ 
\Delta_3(\lambda ) =&- \lambda^3+ \vec{b}_2^2\lambda= -\lambda (\lambda -\sqrt{b_1^2+b_2^2})(\lambda+\sqrt{b_1^2+b_2^2}),  \\ 
\Delta_4(\lambda )  =& \lambda^4 -\vec{b}_3^2\lambda^2 +b_1^2b_3^2= (\lambda^2 - \lambda^2_{4,1})(\lambda - \lambda^2_{4,2}),\\ 
\Delta_5(\lambda )  =&-\lambda^5 +\vec{b}_4^2\lambda^3-(b_1^2b_3^2+b_1^2b_4^2+b_2^2b_4^2)\lambda\\ =& -\lambda (\lambda^2 - \lambda^2_{5,1})(\lambda^2 - \lambda^2_{5,2}),
\end{aligned}
\end{equation} 
where 
\begin{equation}\label{tab2}
\begin{aligned}
\lambda_{4,1}^2 &:=  \frac{1}{2}\left(\vec{b}_3^2 + \sqrt{(\vec{b}_3^2)^2-4b_1^2b_3^2}\right),\\
\lambda_{4,2}^2 &:= \frac{1}{2}\left(\vec{b}_3^2 - \sqrt{(\vec{b}_3^2)^2-4b_1^2b_3^2}\right),\\
\lambda_{5,1}^2 &:= \frac{1}{2}\left(\vec{b}_4^2 + \sqrt{(\vec{b}_4^2)^2-4(b_1^2b_3^2+b_1^2b_4^2+b_2^2b_4^2)}\right),\\
\lambda_{5,2}^2 &:= \frac{1}{2}\left(\vec{b}_4^2 - \sqrt{(\vec{b}_4^2)^2-4(b_1^2b_3^2+b_1^2b_4^2+b_2^2b_4^2)}\right). 
\end{aligned}
\end{equation}
However, one must remember that the matrix elements $b_1, b_2, \ldots, b_L$ in \eqref{qh1} depend on $c_1 = \hbar v_1$ and $c_2 = \hbar v_2 $. An advantage of \eqref{proposition} is that using a well known formulas for the roots of polynomials of degree $2,3$, and $4$, e.g. see \cite{vander}, we can compute explicitly the roots of $\Delta_{L+1}$  for $L\leq 8$. 
From \eqref{matqH} one sees that the spectrum of $\hat{\textbf{H}}$  is the spectrum of $\hat{\textbf{H}}_I$ shifted by $(\omega_1 c_1+\omega_2c_2)$. So, for low dimensional blocks we are able to obtain an explicit formula for reduced Hamiltonian flow $\mathbb{R} \ni t\mapsto e^{\frac{it}{\hbar}\hat{\textbf{H}}}\in \operatorname{Aut} \mathcal{H}_{c_1, c_2}$ and hence, the solutions of the Heisenberg equations \eqref{qa0t}-\eqref{qyt}. 


Now we describe in detail the cases of the lowest dimensional blocks, i.e. $\dim \mathcal{H}_{c_1, c_2} =2$ and $\dim \mathcal{H}_{c_1, c_2} =3$, which will be useful for our further applications. Below we assume $L=\min \{v_1, v_2 \} = v_2$.

\noindent\textbf{Case L=1}. 
Since in this case $v_1 \geq 1$ and $v_2=1$, the basis \eqref{basishc} of $\mathcal{H}_{c_1, c_2}$ contains two elements: $|0, v_1, 1  \rangle$, $|1, v_1-1, 0\rangle$. One has $b_1= g_0\hbar^{\frac{3}{2}}\sqrt{v_1}$  for the coefficient $b_1$  of the matrix \eqref{qh1}. 
The time evolution flow written in this basis is given by 
\begin{equation}\label{evop1}
e^{\frac{i}{\hbar}t\hat{\textbf{H}}} = e^{it(\omega_1 v_1 + \omega_2 )}\begin{pmatrix}
 \cos (\nu t) & i \sin (\nu t)\\
i \sin (\nu t) &  \cos (\nu t)  \end{pmatrix}, 
\end{equation}
where  $\nu := \frac{b_1}{\hbar}= g_0\sqrt{\hbar v_1}$.

Using  \eqref{evop1} we obtain the solution 
\begin{equation}
\textbf{A}_0 (t) = \hbar\begin{pmatrix}
 \sin^2 (\nu t) & -\frac{i}{2} \sin (\nu t)\\
\frac{i}{2} \sin (\nu t) &  \cos^2 (\nu t)  \end{pmatrix}
\end{equation}  
of operator equation \eqref{opeqa0}.

\noindent\textbf{Case L=2}. 
In this case $c_1=\hbar v_1$, $c_2 = 2\hbar $ and $\dim \mathcal{H}_{c_1, c_2} =3$. The subspace $\mathcal{H}_{c_1, c_2}$ is spanned by the vectors \[\{ |0, v_1, 2\rangle , |1, v_1-1, 1 \rangle , |2, v_1-2, 0 \rangle \}.\]
Since $b_1 = g_0\sqrt{\hbar^3 2v_1}$ and $b_2 = g_0\sqrt{\hbar^3 2(v_1-1)}$, the eigenvalues of $\hat{\textbf{H}}_I$ are $0, \hbar \nu, - \hbar \nu$, where 
\begin{equation}\label{defnu}
 \nu := \frac{\sqrt{b_1^2+b_2^2}}{\hbar} =g_0\sqrt{2\hbar (2v_1-1)}.
\end{equation}
We also introduce the quantities:
\begin{equation}\label{betsi}
\beta_1 := \frac{b_1}{\sqrt{b_1^2+b_2^2}} = \sqrt{\left(\frac{1}{2}+\frac{\hbar g_0^2}{\nu^2}\right)} \mbox{ and } \beta_2:= \frac{b_2}{\sqrt{b_1^2+b_2^2}} = \sqrt{\left(\frac{1}{2}-\frac{\hbar g_0^2}{\nu^2}\right)},
\end{equation}
related by $\beta_1^2 + \beta_2^2 = 1$. 
From $v_1 \geq 2$ we immediately obtain that  $\nu \geq g_0\sqrt{6\hbar}$ and $\sqrt{\frac{1}{2}} <\beta_1 \leq \sqrt{\frac{2}{3}}$, $\sqrt{\frac{1}{3}} \leq \beta_2 < \sqrt{\frac{1}{2}}$. 
The time evolution flow in the mentioned above basis is given by 
\begin{multline}\label{evop3}
 e^{\frac{i}{\hbar}t\hat{\textbf{H}}} = e^{it (\omega_1v_1 +2\omega_2)}\\
 \times \begin{pmatrix}
1+ \beta_1^2(\cos (\nu t) -1) & i \beta_1\sin (\nu t) & \beta_1\beta_2 (\cos (\nu t) - 1)\\
i \beta_1\sin (\nu t) & \cos (\nu t) &   i\beta_2\sin (\nu t) \\
\beta_1\beta_2(\cos (\nu t) - 1) & i\beta_2\sin (\nu t) & 1+\beta_2^2( \cos (\nu t)- 1)
\end{pmatrix}.
\end{multline}

Since 
\begin{equation}
\textbf{A}_0 =  \begin{pmatrix}
0& 0 & 0\\
0 & \hbar &   0 \\
0 & 0 & 2\hbar
\end{pmatrix}, 
\end{equation}
the solution $\textbf{A}_0 (t)= e^{\frac{-it}{\hbar}\hat{\textbf{H}}} \textbf{A}_0 e^{\frac{it}{\hbar}\hat{\textbf{H}}}$ of equation \eqref{opeqa0} is the following 
\begin{multline}\label{evop3a0}
 \textbf{A}_0 (t) =  \hbar\sin^2 (\nu t) \begin{pmatrix}
\beta_1^2- 2\beta_1^2\beta_2^2 & 0& \beta_1\beta_2- 2\beta_1\beta_2^3\\
0 & 2\beta_2^2 -1 & 0\\
\beta_1\beta_2- 2\beta_1\beta_2^3 & 0 & \beta_2^2 - 2\beta_2^4
\end{pmatrix} \\
+\hbar\sin (2\nu t) \begin{pmatrix}
0 & -\frac{\beta_1}{2}+2 \beta_1\beta_2^2& 0\\
\frac{\beta_1}{2}-2 \beta_1\beta_2^2 & 0 & \frac{\beta_2}{2}-\beta_2^3\\
0 & -\frac{\beta_2}{2}+\beta_2^3 & 0
\end{pmatrix}\\
+2\hbar\sin (\nu t)\begin{pmatrix}
0 & -\beta_1\beta_2^2& 0\\
\beta_1\beta_2^2 & 0 & -\beta_1^2\beta_2\\
0 & \beta_1^2\beta_2 & 0
\end{pmatrix}\\
+ \hbar\cos (\nu t) \begin{pmatrix}
-4\beta_1^2\beta_2^2 & 0& 2(\beta_1^3\beta_2-\beta_1\beta_2^3)\\
0 & 0 & 0\\
2(\beta_1^3\beta_2-\beta_1\beta_2^3) & 0 & 4b_1^2\beta_2^2
\end{pmatrix}\\
+\hbar\begin{pmatrix}
4\beta_1^2\beta_2^2 & 0& 2(\beta_1\beta_2^3-\beta_1^3\beta_2)\\
0 & 1 & 0\\
2(\beta_1\beta_2^3-\beta_1^3\beta_2) & 0 & 2(\beta_1^4 + \beta_2^4)
\end{pmatrix},
\end{multline}
where $\beta_1$ and $\beta_2$ are defined by \eqref{betsi}.
Having obtained $\textbf{A}_0(t)$, from \eqref{qa0t} one obtains $\textbf{Y}(t)$ and since $\textbf{X}(t)$ is an integral of motion and we have solved the Heisenberg equations \eqref{qa0t}-\eqref{qyt}.


Mathematical model of three non-linearly interacting one-modes given by \eqref{qH} and \eqref{comm} describes in a fully quantum manner various physical phenomena in which photons interact through non-linear media, e.g. through crystals.  There exist a huge number of papers devoted to this subject, see monographs \cite{knight,milburn,perina,boyd} and references therein. In order to describe these non-linear optical phenomena one usually combines the classical description with the quantum one. For example treating one of the modes in the classical way, see \cite{knight,milburn,perina}, allows, after some estimate, to linearize the problem under considerations. Also the description of time evolution of the system has approximate character. Namely, one studies short-time evolution only, e.g. see \cite{perina}.

Completing this section,  we discuss the following mutually inverse processes: 
\begin{enumerate}
\item[(i)] parametric up-conversion ( sum-frequency generation)
\begin{equation}
| 0, v_1, 2 \rangle \mapsto |2, v_1-2, 0 \rangle 
\end{equation}
\item[(ii)] parametric down-conversion (difference-frequency generation) 
\begin{equation}
|2, v_1-2, 0\rangle \mapsto |0, v_1, 2\rangle ,
\end{equation}
\end{enumerate}
where we have assumed $\omega_0 = \omega_1 + \omega_2 $ and $L=2$.  For simplicity we will avoid the use of the `historical' terminology: pump, signal and idler, for the considered modes. Therefore, in the case (i) one converts two photons of frequencies $\omega_1$ and $\omega_2 $ of modes $1$ and $2$,respectively into two photons of frequency $\omega_1 + \omega_2$ of $0$-mode. In the case (ii) the two photons of $0$-mode of frequency $\omega_0$ convert to two pairs of photons of frequencies $\omega_1$ and $\omega_2$ of $1$ and $2$ modes, respectively.

The transition probability for these processes is the same and its dependence on time is given by 
\begin{multline}\label{probtrans}
|\langle 0, v_1, 2 | e^{\frac{it}{\hbar} \hat{\textbf{H}}} | 2, v_1-2, 0 \rangle |^2 = \left(\frac{1}{4}- \left(\frac{\hbar g_0^2}{\nu^2}\right)^2\right) (\cos (\nu t) - 1)^2. 
\end{multline}
So, it oscillates in time with frequency and amplitude dependent on the number of photons in $1$-mode. As follows from \eqref{defnu}, with increasing number of photons $v_1$ in $1$-mode  the frequency of oscillations is growing and the amplitude tends to $\frac{1}{4}$. Moreover, as one can see from \eqref{probtrans}, at time  $t= \frac{2K\pi }{\nu}$, $K\in \mathbb{N}$ the probability of transition between states in the considered processes is zero.

Some class of non-linear two-mode systems were investigated in \cite{gol2}, where authors discussed their integrability and applications in optics.

\section{Reduced coherent states}\label{sec4}

In this section, combining the classical and quantum reduction procedures, we will obtain the reduced coherent states as a result of the general construction presented in \cite{KS}. Here by the reduced coherent state map we understand a map $\mathcal{K}_{c_1, c_2}: \mathcal{C}^{-1}(0)\backslash \{ \vec{0}, \vec{c} \} \to \mathbb{C}\mathbb{P} (\mathcal{H}_{c_1, c_2})$ of the classical reduced phase space $\mathcal{C}^{-1}(0)\backslash \{ \vec{0}, \vec{c}\}$ into the quantum reduced phase space $\mathbb{C}\mathbb{P} (\mathcal{H}_{c_1, c_2})$. We will also derive an explicit formula for the reproducing measure for the reduced coherent states.

Since Hilbert space $\mathcal{H}$ can be decomposed into the subspaces $\mathcal{H}_{c_1, c_2}$ 
\begin{equation}
\mathcal{H} = \bigoplus_{(c_1, c_2) \in \hbar\mathbb{Z}_+^2 } \mathcal{H}_{c_1, c_2}, 
\end{equation}
where $\mathbb{Z}_+^2 := \{ (n_1, n_2) \in \mathbb{Z} : n_1 \geq 0, n_2 \geq 0 \}$, we can also decompose the standard coherent states:
\begin{equation}
|z_0, z_1, z_2 \rangle = \sum_{(c_1, c_2)\in \hbar\mathbb{Z}_+^2} P_{c_1, c_2} |z_0, z_1, z_2 \rangle, 
\end{equation}
where $P_{c_1, c_2}$ is the orthogonal projection of $\mathcal{H}$ on Hilbert subspace $\mathcal{H}_{c_1, c_2}$. For the chosen subspace $\mathcal{H}_{c_1, c_2}$ one has 
\begin{equation}\label{zfactorem}
P_{c_1, c_2} |z_0, z_1, z_2\rangle  = \frac{z_1^{v_1}z_2^{v_2}}{\sqrt{\hbar^{v_1+v_2}}}K_{c_1, c_2} (\hat{z}),
\end{equation}
where $K_{c_1, c_2}: \mathbb{C} \to \mathcal{H}_{c_1, c_2}$ is a map defined by 
\begin{equation}\label{redcohstates}
K_{c_1, c_2} (\hat{z}) = |\hat{z}; c_1, c_2 \rangle := \sum_{n=0}^L \frac{\hat{z}^n}{g_0^n \sqrt{n! (v_1-n)!(v_2-n)!}} \hbar^{\frac{n}{2}} |n, v_1-n, v_2-n \rangle . 
\end{equation}
and the complex variable $\hat{z}$ is related to $(z, I_0)\in \mathcal{C}^{-1}(0)\backslash \{\vec{0}, \vec{c} \}$ by
\begin{equation}
\label{cor41}
\hat{z} = \frac{1}{|z_1|^2 |z_2|^2 } z = \frac{1}{(c_1 - I_0)(c_2-I_0)} z = g_0\sqrt{\frac{I_0}{(c_1-I_0)(c_2 - I_0)}} e^{i\psi_0}.
\end{equation}
The last equality in \eqref{cor41} follows from \eqref{variablezet}. Using \eqref{zfactorem} and \eqref{cor41} one obtains the smooth injective map $\mathcal{K}_{c_1, c_2} : ]0, c[ \times \mathbb{S}^1 \to \mathbb{C}\mathbb{P} (\mathcal{H}_{c_1, c_2} )$ from the reduced phase space $]0, c[ \times \mathbb{S}^1 \cong \mathcal{C}^{-1}(0)\backslash \{ \vec{0}, \vec{c}\}$ into $\mathbb{C}\mathbb{P} (\mathcal{H}_{c_1, c_2} )$. The complex coefficient which appeared on the right hand side of \eqref{zfactorem} written in the canonical coordinates $(I_0, I_1, I_2, \psi_0, \psi_1, \psi_2)$ takes the form
\begin{equation}
 \frac{z_1^{v_1}z_2^{v_2}}{\sqrt{\hbar^{v_1+v_2}}} = \left(v_1 - \frac{I_0}{\hbar}\right)^{\frac{v_1}{2}}\left(v_2 - \frac{I_0}{\hbar}\right)^{\frac{v_2}{2}} e^{i(v_1 \psi_1 + v_2 \psi_2)}. 
\end{equation}

For the reduced coherent states \eqref{redcohstates} we have the following resolution of identity
\begin{equation}\label{resofidentity}
P_{c_1, c_2} \mathbbm{1} = \int_{\mathbbm{C}\backslash \{0\}} |z; c_1, c_2 \rangle \langle z; c_1, c_2 | d\nu_{c_1, c_2} (\bar{\hat{z}}, \hat{z}).
\end{equation}
In order to find the weight function $\rho$ defining the measure $d\nu_{c_1, c_2} (\overline{\hat{z}}, \hat{z}) = \rho ( |\hat{z}|^2) d|\hat{z}|^2 d\psi$ appearing in \eqref{resofidentity}, where $\hat{z} = |\hat{z}|e^{i\psi}$, we use the integral formula presented  in \cite[Proposition 5.1. eq. (5.9)]{KS}, which in this particular case takes the form
\begin{equation}
\rho (|\hat{z}|^2) = \frac{1}{2\pi \hbar^{3+v_1+v_2} g_0^2} \int_0^\infty \int_0^\infty x_1^{v_1+1} x_2^{v_2+1} e^{-\frac{1}{\hbar}(\frac{|\hat{z}|^2 x_1 x_2}{g_0^2} +x_1 + x_2)} dx_1 dx_2. 
\end{equation}
Calculating integrals on the right hand side of the above equation we obtain the explicit formula for the weight function $\rho$:
\begin{equation}\label{rofun}
\rho (x) = \frac{(v_1+1)!(v_2+1)!}{2\pi \hbar^{\frac{v_1+v_2+1}{2}}} g_0^{v_2+v_1+1} e^{\frac{g_0^2}{2\hbar x}} W_{-\frac{(v_1+v_2+3)}{2} ; \frac{v_1-v_2}{2}}\left(\frac{g_0^2}{\hbar x}\right) , 
\end{equation}
where $ W_{-\frac{(v_1+v_2+3)}{2} ; \frac{v_1-v_2}{2}}$ is the Whittaker function (see \cite{Ryzhik} for its definition). 

One can easily rewrite the resolution of identity \eqref{resofidentity} as reproducing property 
\begin{equation}
 \psi (w) = \int_{\mathbbm{C}\backslash \{0\}} \psi (z) \langle \hat{z}; c_1, c_2 |\hat{w}; c_1, c_2 \rangle  d\nu_{c_1, c_2} (\bar{\hat{z}}, \hat{z})
\end{equation}
for a kernel
\begin{multline}
\langle \hat{z}; c_1, c_2 | \hat{w}; c_1, c_2 \rangle = \sum_{n=0}^L \frac{(\bar{\hat{z}}\hat{w}\hbar)^n }{g_0^{2n} n!(v_1-n)!(v_2-n)!} =\\
= \frac{1}{v_1!v_2!} \mbox{ } _2F_0\left[\begin{matrix}
-v_1, & -v_2 \\
 & - 
\end{matrix}; \frac{\bar{\hat{z}} \hat{w} \hbar }{g_0^2}\right], 
\end{multline}
where 
\begin{equation}\label{psizeta}
\psi (\hat{z}) := \langle \psi | \hat{z}; c_1, c_2 \rangle 
\end{equation}
for $|\psi \rangle \in \mathcal{H}_{c_1, c_2}$.  Formula \eqref{psizeta} defines an antilinear isomorphism $|\psi \rangle \mapsto \psi (z) $ of $\mathcal{H}_{c_1, c_2} $ with the space $L^2 (\mathbb{C}\backslash \{0\} , d\nu_{c_1, c_2} )$ of polynomials  on $\mathbb{C}\backslash \{0\}$ of degree less or equal than $L$, where the scalar product of $\psi, \phi \in L^2 (\mathbb{C}\backslash \{0\} , d\nu_{c_1, c_2} )$ is defined by 
\begin{equation}
\langle \psi |\phi \rangle : = \int_{\mathbbm{C}\backslash \{0\}} \overline{\psi (\hat{z})} \phi (\hat{z}) d\nu_{c_1, c_2} (\bar{\hat{z}}, \hat{z}) . 
\end{equation}


Using \eqref{a0nabaze} -\eqref{azgnabaze} and \eqref{redcohstates} we find that $\textbf{A}_0, \textbf{A}$ and $\textbf{A}^*$ act on the reduced coherent states as follows 
\begin{equation}\label{rednastanyred}
\begin{aligned}
\textbf{A}_0 |\hat{z}; c_1, c_2 \rangle & = \hbar \hat{z} \frac{d}{d\hat{z}} |\hat{z}; c_1, c_2 \rangle , \\
\textbf{A}|\hat{z}: c_1, c_2 \rangle & = \hat{z} (c_1 - \hbar \hat{z} \frac{d}{d\hat{z}})(c_2 - \hbar\hat{z} \frac{d}{d\hat{z}}) |\hat{z}; c_1, c_2 \rangle , \\
\textbf{A}^* |\hat{z} ; c_1, c_2 \rangle & = g_0^2 \hbar \frac{d}{d\hat{z}}|\hat{z}; c_1, c_2 \rangle. 
\end{aligned}
\end{equation}
Hence, using isomorphism \eqref{psizeta}  one finds from \eqref{rednastanyred} that $\textbf{A}_0, \textbf{A}$ and $\textbf{A}^*$ are represented in $L^2 (\mathbb{C}\backslash \{0\} , d\nu_{c_1, c_2} )$ as the differential operators
\begin{equation}\label{rednafunkcje}
\begin{aligned}
\textbf{A}_0 \psi (\hat{z}) & = \hbar \hat{z} \frac{d}{d\hat{z}} \psi (\hat{z}), \\
\textbf{A} \psi (\hat{z}) & = g_0^2 \hbar \frac{d}{d\hat{z}} \psi (\hat{z}), \\ 
\textbf{A}^* \psi (\hat{z})& = \hat{z} (c_1 - \hbar \hat{z} \frac{d}{d\hat{z}})(c_2 - \hbar\hat{z} \frac{d}{d\hat{z}}) \psi (\hat{z} ). 
\end{aligned}
\end{equation}
So, we can formulate the eigenproblem $\hat{\textbf{H}} |\psi \rangle = \lambda |\psi \rangle $ for the reduced Hamiltonian \eqref{redqH} in the form of differential equation 
\begin{multline}\label{diffeqnalambda}
 z^3 \frac{d^2}{dz^2}\psi (z) + \left((1-v_1+v_2) z^2 +\frac{1}{\hbar}(\omega_0 - \omega_1 - \omega_2) z  + \frac{g_0^2}{\hbar}\right) \frac{d}{dz}\psi (z) \\
+ (v_1v_2z + \frac{1}{\hbar}(\omega_1v_1 + \omega_2 v_2)) \psi (z) = \lambda \psi (z),
\end{multline}
where $\lambda \in \mathbb{R}$ is spectral parameter.

\end{document}